# Numerical investigation of progressive damage and associated seismicity on a laboratory fault

Qi Zhao[1,2*], Nicola Tisato[3], Aly Abdelaziz[2], Johnson Ha[2], and Giovanni Grasselli[2]

[1]Department of Civil and Environmental Engineering, The Hong Kong Polytechnic University, Hung Hom, Hong Kong SAR, China.

[2]Department of Civil and Mineral Engineering, The University of Toronto, 35 St. George Street, Toronto, Ontario M5S 1A4, Canada.

[3]Department of Geological Sciences, Jackson School of Geosciences, The University of Texas at Austin, 2305 Speedway Stop C1160, Austin, TX 78712-1692, USA.

*Corresponding author: Qi Zhao (qi.qz.zhao@polyu.edu.hk)

**Abstract**

Understanding rock shear failure behavior is crucial to gain insights into slip-related geohazards such as rock avalanches, landslides, and earthquakes. However, descriptions of the progressive damage on the shear surface are still incomplete or ambiguous. In this study, we use the hybrid finite-discrete element method (FDEM) to simulate a shear experiment and obtain a detailed comprehension of shear induced progressive damage and the associated seismic activity. We built a laboratory fault model from high resolution surface scans and micro-CT imaging. Our results show that under quasi-static shear loading, the fault surface experiences local dynamic seismic activities. We found that the seismic activity is related to the stress concentration on interlocking asperities. This interlocking behavior (i) causes stress concentration at the region of contact that could reach the compressive strength, and (ii) produces tensile stress up to the tensile strength in the region adjacent to the contact area. Thus, different failure mechanisms and damage patterns including crushing and sub-vertical fracturing are observed on the rough surface. Asperity failure creates rapid local slips resulting in significant stress perturbations that alter the overall stress condition and may trigger the slip of adjacent critically stressed asperities. We found that the spatial distribution of the damaged asperities and the seismic activity is highly heterogeneous; regions with intense asperity interactions formed gouge material, while others exhibit minimal to no damage. These results emphasize the important role of surface roughness in controlling the overall shear behavior and the local dynamic seismic activities on faults.

# 1 Introduction

Understanding shear behavior along rock discontinuities at various scales, such as joints and faults, is essential to rock engineering projects and geohazard mitigation. Rock discontinuities are planes of weakness and are responsible for many geohazards, for example, rock avalanches, landslides, and earthquakes. Numerous laboratory shear experiments have been conducted on a large variety of rock types under different conditions (e.g., Bandis et al., 1983; Beeler, 1996; Marone, 1998; Di Toro et al., 2004; Grasselli, 2001; Reches and Lockner, 2010; Tisato et al., 2012; Kim and Jeon, 2019; Zhao et al., 2020; Morad et al., 2022). Among these studies, many suggested the importance of surface roughness and contact condition in controlling the shear behavior. However, the progressive damaging process on faults is still not well understood because fault surfaces cannot be observed directly during shear, with the exception of a few studies utilizing transparent halite samples (e.g., Renard et al., 2012) and in situ and operando testes conducted under X-ray micro-computed tomography (micro-CT) (e.g., Zhao et al, 2018; Zhao et al., 2020). Shear processes control coseismic damage and friction on fault, but the constitutive friction theories are not yet fully understood.

To observe and gain insights into damage processes on rough rock surfaces undergoing shear deformation, Tatone and Grasselli (2013) used micro-CT to image the joint surfaces after the shear test; and Crandall et al. (2017) used micro-CT to obtain geometrical information from fractured shale core that is incrementally sheared. Recently, direct and detailed observations of the evolution of laboratory fault were achieved by using an *in situ* rotary shear experimental apparatus under X-ray micro-CT (Zhao et al., 2017). Such experimental results help draw the connections between microscopic damage and macroscopic shear behavior, the formation and accumulation of gouge material, and shear-induced secondary fractures (Zhao, 2017; Zhao et al., 2018; Zhao et al., 2020). However, due to technological limitations, time-continuous observations of the shear surface evolution and the *in situ* stress condition remain shortfalls.

Numerical simulation methods have been extensively used to study the shear behavior of rock discontinuities. To simulate the interaction and breakage of asperities and the frictional sliding behavior, numerical methods that can capture solid fracturing and interaction are typically used. For example, the particle-based lattice solid model (Mora and Place 1993) uses a numerical concept similar to the discrete element method (DEM) to simulate frictional behavior and fracturing in solids. Mora and Place (1998) and Place and Mora (2000) used their model to study the role fault gouge on the frictional behavior of faults, offering a possible explanation for the heat flow paradox (Henyey and Wasserburg, 1971; Lachenbruch and Sass, 1992). Bonded particle-based methods, such as the particle flow code (PFC) (Cundall and Strack, 1979), are commonly used for simulating rock shear behavior. Park and Song (2009) used PFC3D to simulate direct shear tests and demonstrated that the method can simulate typical rock joint shear behavior, and they found that the peak shear strength and peak dilation angle was strongly influenced by the friction coefficient, roughness, and bond strength, while the residual shear strength and residual friction angle was influenced by the particle size, friction coefficient, and bond strength. Asadi et al. (2012) used a similar approach in two dimensions (PFC2D) to simulate direct shear experiments on synthetic joint profiles of varying roughness and boundary conditions to assess asperity degradation and intact material damage. They showed that as the joint is sheared, highly localized asperity interaction on the joint surface and the geometry of the asperities has a significant influence on how joints fail. Types of failure include asperity sliding, cut-off, separation, and crushing, typically associated with tensile failure into the intact material in conditions with steep asperities and/or high normal stress. However, Bahaaddini et al. (2013) identified a significant shortcoming of the particle-based methods due to the unrealistic shear and dilation behavior of joints as a result of particle interlocking due to the inherent micro-scale roughness of the joint. To overcome this limitation, they implemented the smooth-joint model (Pierce et al., 2007), where the blocks associated with either side of the joint are generated separately to appropriately define and apply the smooth joint model to the interface. Lambert and Coll (2014) created a synthetic rock joint by importing the real morphology of the joint surface into a bonded particle assembly and studied the shear behavior using the smooth-joint model. Their results reproduced the progressive degradation of the asperities upon shearing.

The hybrid finite-discrete element method (FDEM) is increasingly used to investigate rock shear behavior. FDEM is a micromechanical numerical method first introduced by Munjiza et al., (1995) combining the finite element method and discrete element method. In doing so, the numerical method can model the transition of a continuous material to a discontinuous material as it deforms, yields, and breaks. Karami and Stead (2008) and Tatone (2014) used FDEM to model direct shear tests and relate progressive asperity degradation mechanisms with the measured shear stress and dilation during shearing. In addition, Tatone (2014) verified the numerical modelling observations by coupling their study with X-ray micro-CT imaging on post-mortem specimens. It was found that tensile fractures develop in asperities at and near the peak shear stress, followed by a reduction in shear resistance as asperities continue to fail in both tension and shear, and finally, a residual shear resistance is reached once asperities are completely broken and gouge is formed.

In this study, we used the two-dimensional (2D) FDEM to simulate an experiment on the gradual evolution of deformation of a laboratory fault, and we improve the understanding of shear behavior of rough faults through combined interpretation of the simulation and experimental results. First, we provide a brief review of the FDEM, emphasizing the modeling of damage and seismic activity. Second, we develop a clustering algorithm to improve the comprehension of the simulated fractures and seismic events. Next, we build a model based on the laboratory experiment and analyze the simulated results focusing on three aspects that are hardly accessible by experiments: (1) the time-continuous variation of stress conditions on the shear surface, (2) the progressive failure of the asperities and accumulation of gouge, and (3) the seismic activity related to shear-induced damage.

The carefully built and calibrated numerical model is able to simulate the emergent rock mechanical and frictional behaviors. We observe that shear-induced damage and seismic activities are heterogeneously distributed along the fault surface due to the surface roughness. Seismic events occur at the locations of asperity failure due to the interlocking-induced stress concentration. Such events radiate seismic waves and significantly change the overall stress conditions. Some areas on the fault were covered by gouge material and free from damage. These results agree with the laboratory observations and further elaborate on the importance of surface roughness in controlling shear behavior, which is critical to rock engineering practices and earthquake studies.

## 2 Material and methods

### 2.1 In situ shear test under X-ray micro-CT

The numerical simulation in this study is based on the experimental work using *in situ* shear tests under micro-CT reported by Zhao et al. (2018) and Zhao et al. (2020), and a brief review is provided here for completeness. The tested specimen was a cylindrical Flowstone (microfine calcium sulfate cement mortar) 32 mm in length and 12 mm in diameter. The specimen was divided into top and bottom parts by a three-point bending test that created two semi-samples divided by a discontinuity (i.e., laboratory fault) with two matching rough surfaces. An unconfined rotary shear test was conducted on the two semi-samples by shearing the fault under the initial normal stress of 2.5 MPa. The top semi-sample was forced to slip incrementally against the fixed bottom semi-sample. Normal force and torque were recorded during rotation and used to calculate the friction coefficient. After each incremental slip of 6°, a three-dimensional (3D) micro-CT scan was conducted, which allows for imaging of the gradual morphological evolution of the specimen (Fig. 1a). This experimental work provided detailed information of the shear-induced secondary fractures (Fig. 1b) and the progressive damage on the slipping surface (Fig. 1c) in the sample volume; however, the observation of the shear surface damage evolution was only available at discrete time points coincident with each shear step, while an actual time-continuous observation of the shear surface evolution and the local stress condition on the rough surface was not available.

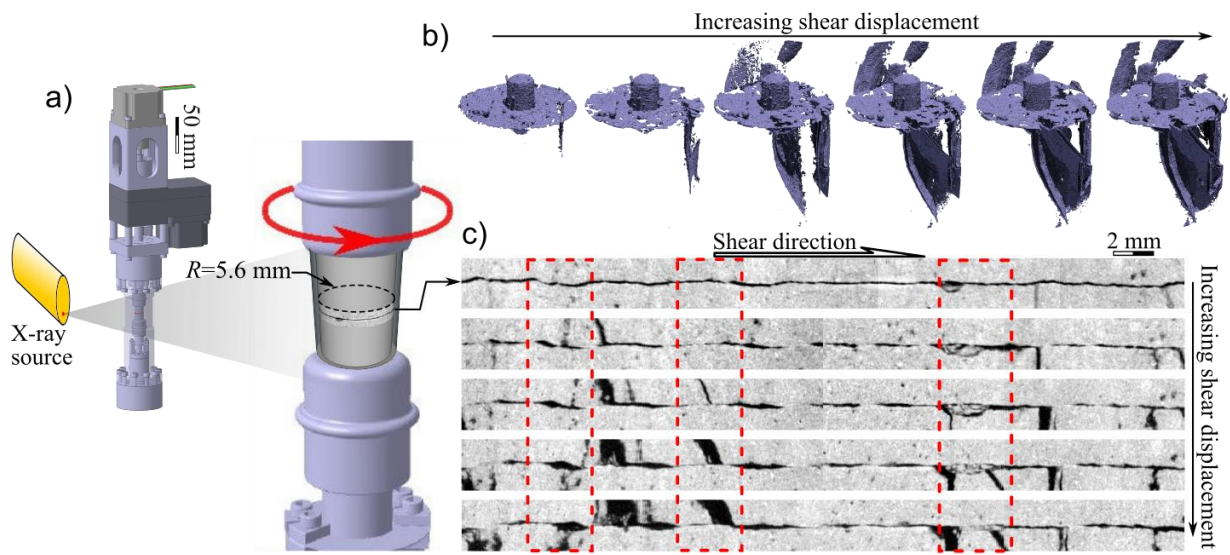

**Fig. 1** Summary of the laboratory set-up and results. (a) Schematic of the in situ shear test and a zoom-in view of the shear surface (i.e., the zone of interest). (b) 3D visualization of the development of shear induced fractures with increasing shear displacement. (c) 2D unwrapped micro-CT image slice showing the progressive damage on the slipping surface with increasing shear displacement, viewed at the radius ($R$ = 5.6 mm) corresponding to the highest asperity. Red

dashed boxes in (c) indicate (from left to right) shear-induced aperture opening, fracturing, and progressive damage and gouge formation (modified from Zhao et al. (2018) and Zhao et al. (2020)).

## 2.2 The hybrid finite-discrete element method

The hybrid finite-discrete element method (FDEM) combines continuum mechanics principles with discrete element principles to simulate interaction, deformation, and fracturing of materials (Munjiza et al., 1995; Munjiza, 2004). FDEM has been used to investigate a wide range of rock mechanics and geophysics problems including, but not limited to, tunneling and excavation, micromechanics, rock joint shear behavior, hydraulic fracturing, thermal-mechanical/hydro-thermal-mechanical coupling, and fault dynamics (e.g., Karami and Stead 2008; Mahabadi et al., 2012; Lisjak et al., 2014; Zhao et al., 2014; Yan et al., 2016; Huang et al., 2017; Lei et al., 2017; Ma et al., 2017; Fukuda et al., 2019; Okubo et al., 2019; Knight et al., 2020). Simulating the entire shear behavior and evolution of a rough surface is a challenging task that requires advanced computational resources encompassing, for example, the 3D FDEM method. However, 3D models explicitly capturing the surface roughness at sub-millimeter resolution and the entire shear process is not practical due to the demanded computation power. On the other hand, 2D FDEM simulations has the merit of reducing the computational demand, and it has been shown to provide insights into the mechanical behavior of rock joints and faults (e.g., Karami and Stead, 2008; Tatone, 2014; Okubo et al., 2019).

FDEM models synthesize the macroscopic behavior of materials from the interaction of the micromechanical constituents. In a 2D FDEM model, the simulated material is first discretized based on a finite element mesh consisting of nodes and triangular elements. Then, the finite element mesh is enriched by inserting a four-node cohesive crack element (CCE) between each adjacent triangular element pair. Motion for the discretized system is calculated by an explicit time integration scheme, and the nodal coordinates of the elements are updated at each simulation step (Munjiza, 2004). FDEM models the progressive damage and failure of brittle material according to the principles of non-linear elastic fracture mechanics (Dugdale, 1960; Barenblatt, 1962), and it captures the fracturing behavior of solids by modeling the entire failure path, including elastic deformation, yielding, and fracturing (Fig. 2).

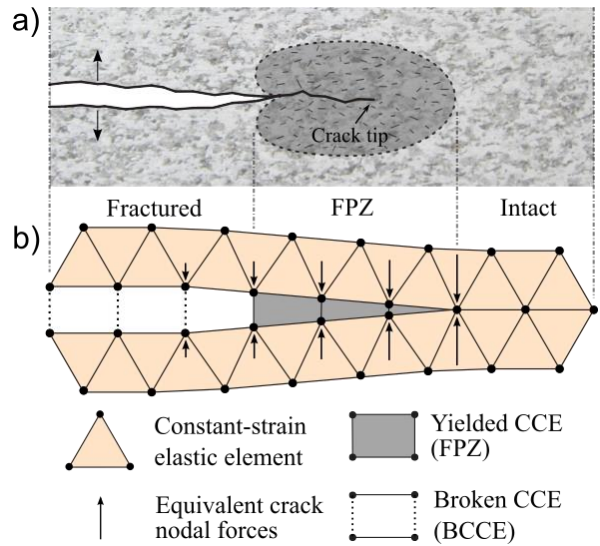

**Fig. 2** Schematic diagram showing the FDEM approach of simulating fracturing. (a) Propagation of a fracture and the creation of fracture process zone (FPZ). (b) Realization of the fracturing process in FDEM involves the yielded cohesive crack elements and broken cohesive crack elements (BCCE).

Depending on the local stress and deformation field, the CCE undergoes elastic deformation, yielding, and breakage, simulating the damage development of the fracture process zone (FPZ) (Fig. 3) (Labuz et al., 1985). During elastic loading, the relationships between bonding stresses (normal bonding stress, $\sigma$ and shear bonding stress, $\tau$) and the corresponding crack displacement (opening, $o$ and slip, $s$) are as follows (Munjiza et al., 1999):

$$\sigma = \begin{cases} \frac{2o}{o_p} f_t & (o < 0, \text{compression}) \\ \left[ \frac{2o}{o_p} - \left( \frac{o}{o_p} \right)^2 \right] f_t & (0 < o < o_p, \text{tension}) \end{cases} \tag{1}$$

$$\tau = \left[ \frac{2s}{s_p} - \left( \frac{s}{s_p} \right)^2 \right] f_s \quad (|s| \leq |s_p|, \text{shear}) \tag{2}$$

where $f_t$ and $f_s$ are the peak tensile and shear bonding strength of a CCE, respectively. The peak shear bonding strength is calculated based on the Mohr-Coulomb failure criterion using the cohesion ($c$) and internal friction angle ($\phi$): $f_s = c + \sigma \tan \phi$. $o_p$ and $s_p$ are the peak opening and slip values at the peak bonding stresses calculated as $o_p = 2hf_t/p_f$ and $s_p = 2hf_s/p_f$, where $h$ is the nominal element edge length, and $p_f$ is the fracture penalty value. A CCE yields once the stress reaches the peak, then it experiences a post-peak softening behavior with the bonding stresses gradually decreased (Munjiza et al., 1999):

$$\sigma = F(D)f_t \tag{3}$$

$$\tau = F(D)f_s \tag{4}$$

$F(D)$ is an empirical function that approximates the shape of the experimental stress-displacement failure curve according to Evans and Marathe (1968):

$$\text{F}(D) = \left[ 1 - \frac{a+b-1}{a+b} \exp\left( D \frac{a+cb}{(a+b)(1-a-b)} \right) \right] \cdot \left[ a(1-D) + b(1-D)^c \right] \tag{5}$$

where a, b, c are empirical curve fitting parameters equal to 0.63, 1.8, and 6.0, respectively. The damage coefficient ($D$) is calculated for Mode I, II, and I-II as

$$D_{\text{I}} = \frac{o - o_p}{o_r - o_p} \tag{6}$$

$$D_{\text{II}} = \frac{s - s_p}{s_r - s_p} \tag{7}$$

$$D_{\text{I-II}} = \sqrt{D_{\text{I}}^2 + D_{\text{II}}^2} \qquad (8)$$

with the subscripts indicating the mode of failure. The CCE breaks when $D = 1$, which corresponds to a residual opening ($o_{\text{r}}$) or a residual slip ($s_{\text{r}}$), for pure Mode I or II failure, respectively. For Mode I-II failure, $D_{\text{I-II}} = 1$ corresponds to a mixed failure opening and slip ($o_{\text{f}}$ and $s_{\text{f}}$). The values of $o_{\text{r}}$ and $s_{\text{r}}$ are calculated using the predefined numerical fracture energy $G^{\text{f}}_{\text{I}}$ and $G^{\text{f}}_{\text{II}}$, for opening failure and shear failure, respectively. The failure mode of the CCE ($\kappa$) is computed as

$$\kappa = \begin{cases} 1 & \text{(pure tensile, Mode I)} \\ 1 + D_{\text{II}} & \text{(mixed mode, Mode I} - \text{II)} \\ 2 & \text{(pure shear, Mode II)} \end{cases} \qquad (9)$$

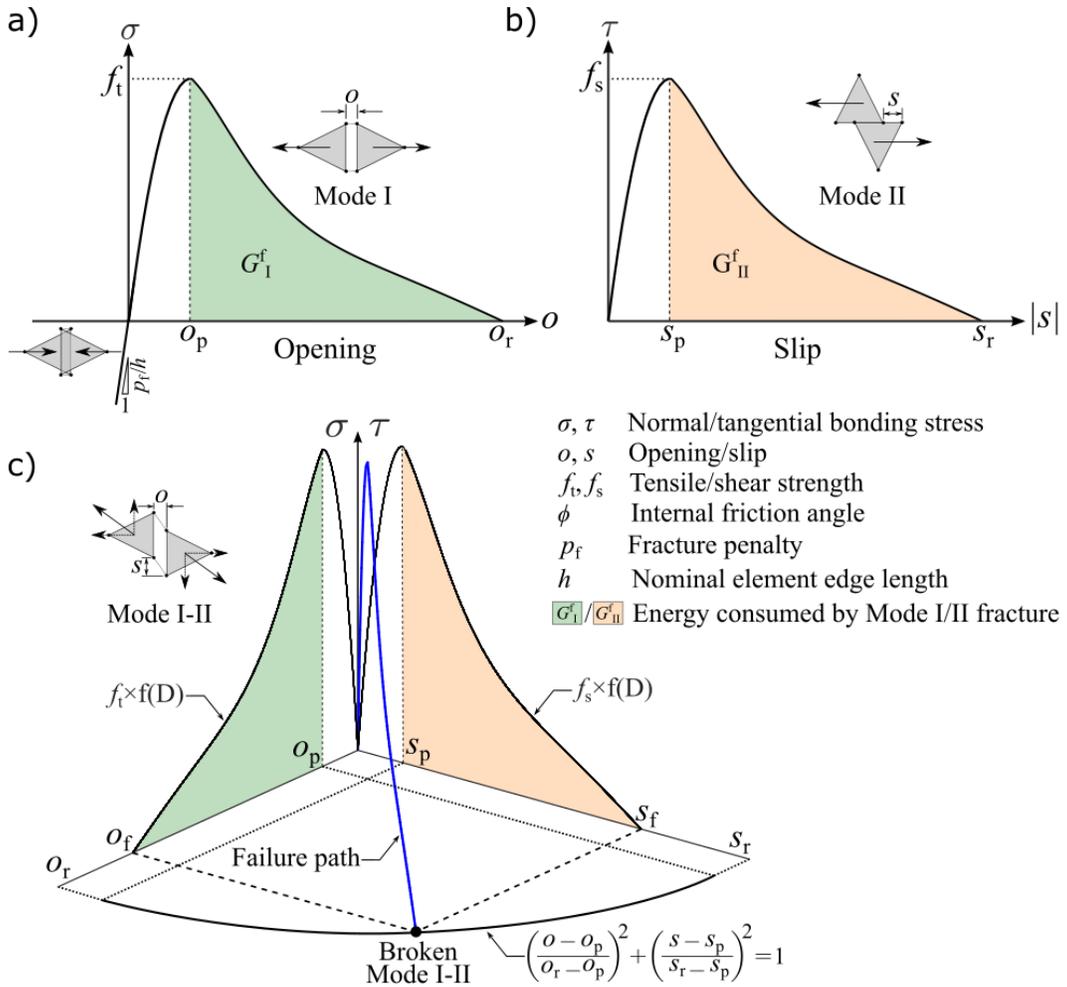

**Fig. 3** Deformation and failure criteria of the cohesive crack element (CCE). (a) Mode I, tensile mode, (b) Mode II, shear mode, and (c) Mode I-II, mixed-mode. Shaded areas highlight the total fracture energy consumed during the failure process of a CCE. The blue curve (failure path) indicates the stress condition during the yielding and failure processes of the CCE.

When both $D_I$ and $D_{II}$ are satisfied at the same time, the failure is also considered as Mode I-II, and a value of 1.5 is assigned to these events during post-processing. The broken cohesive crack element (BCCE) is then considered as a new crack with no cohesion, and its behavior is handled by the interaction algorithms, which are discussed in detail in the literature (Munjiza 2004; Mahabadi et al., 2012).

### 2.3 Simulation of fracture propagation and seismicity in FDEM

Modeling seismic activity in rocks can provide quantitative information of the rock failure process, and a validated model can improve the understanding of laboratory and field seismic observations. In FDEM, upon breakage of the CCE, the accumulated strain energy is released, resembling seismic activity. The coordinates, failure time, kinetic energy at failure, and failure mode of the related BCCE can be recorded (Lisjak et al., 2013). However, a limitation of this approach is that it considers each BCCE as one single seismic event. Consequently, the properties of the fracture and the associated seismic events are highly dependent on the mesh size and mesh orientation (Munjiza and John, 2002). In nature, the breakage of CCEs can be regarded as acoustic emissions associated with the breakage of several mineral grains and grain boundaries (Zhao et al., 2015; Abdelaziz et al., 2018). In most cases, such a mesh dependency needs to be addressed to obtain a better physical meaning of the failure process of CCEs. Zhao et al. (2014) attempted to mitigate the problem with a clustering algorithm considering the temporal and spatial distribution of BCCEs. In their method, each BCCE is viewed as an advancing crack tip, and BCCEs connecting to the crack tip are clustered together as a continuous fracture. However, this implementation did not consider the physical meaning of fracture propagation. The propagating fracture can arrest and then continue to propagate according to the stress conditions and material heterogeneities (Van der Pluijm and Marshak, 2004), and from an energy dissipation point of view, choosing the yielding point of a BCCE as the fracture tip is more consistent with the cohesive crack model (Shet and Chandra, 2002).

Stemming from Zhao et al. (2014) and Zhao (2017), we implemented a new clustering algorithm to mimic fracture propagation process during a seismic event. Note that we consider only seismic activities related to the formation of new fractures, and seismic events created by slipping on existing fracture surfaces are not considered. The algorithm proceeds as follows:

(1) The first BCCE that yields at time $t_y$ and fails at time $t_f$ is considered the initial crack of a cluster. The search algorithm is then executed to include BCCEs connecting to either side of this BCCE (i.e., fracture tips).

(2) BCCEs that are connected to the fracture tips and yield within the time window between $t_y$ and $t_f$ are included in the same cluster and then treated as new fracture tips. At each output frame, the same searching criterion is applied to such new fracture tips until no new BCCEs are found. Then, this cluster of BCCEs is considered to be one continuous fracture, whose growth has produced one seismic event.

(3) Repeat steps 1-2, until all recorded BCCEs are processed.

(4) Calculate the source parameters of the clustered seismic events as follows (for a cluster of $n$ BCCEs):

(a) event time is the breakage time $t_f$ of the initial BCCE in this cluster;

(b) the hypocentre location is the centre coordinates of the initial BCCE in this cluster;

(c) the kinetic energy, $E_e$, is calculated as the sum of the kinetic energy of all BCCEs in this cluster, $E_e = \sum_{i=1}^{n} E_k^i$, where $E_k^i = \frac{1}{2} \sum_{j=1}^{4} m_j v_j^2$ is the kinetic energy of a BCCE, and $m_j$ and $v_j$ are the nodal mass and velocity of the BCCE at the time of breakage. We adopt the empirical relation between radiated energy and magnitude to calculate the magnitude of the seismic events: $M_e = \frac{2}{3}(\log E_e - 4.8)$ (Gutenberg, 1956; Lisjak et al., 2013).

(d) the dominant source mechanism ($\zeta$) of each cluster is calculated as a weighted average of the failure modes of all BCCEs in this cluster:

$$\zeta = \frac{\sum_{i=1}^{n} E_k^i \kappa^i}{\sum_{i=1}^{n} E_k^i} \tag{10}$$

Where $\kappa^i$ is the failure mode of the $i$th BCCE, and its associated kinetic energy, $E_k^i$ is taken as its weight. $\zeta = 1$ and 2 represent pure tensile (Mode I) and shear events (Mode II), respectively, while events having $1 < \zeta < 2$ have tensile and shear failure components (Mode I-II).

This algorithm considers multiple BCCEs created by a single fracturing event, resulting in a more realistic representation of the source mechanism and event energy than previous studies. Note that if a series of connected CCEs break simultaneously due to mechanisms such as crushing or pulverization, they will also be clustered as one event under this algorithm.

## 2.4 Numerical model setup

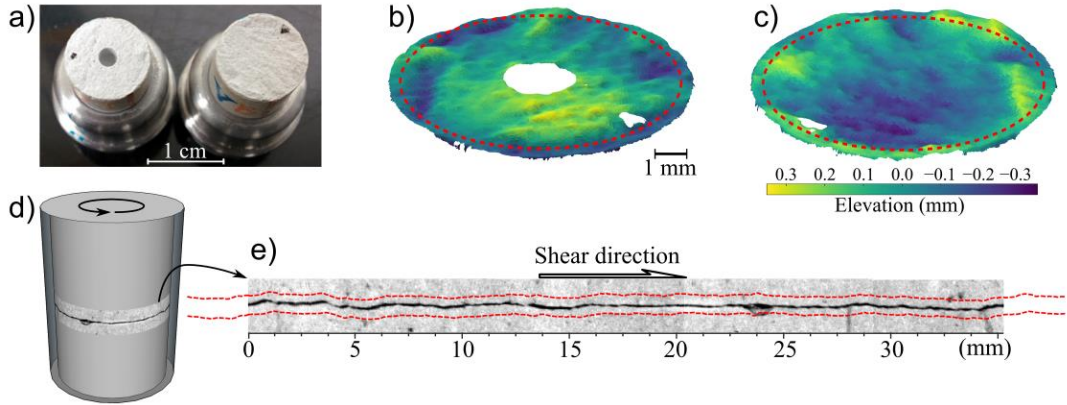

**Fig. 4** Preparation of 2D surface profiles for the FDEM model. (a) The top (left) and bottom (right) parts of the sample used in the rotary shear experiment. (b)–(c) 3D surface scan of the shear surfaces. Red dashed lines indicate the extracted profiles. (d) The initial condition by micro-CT imaging. (e) Comparison of the profiles (red dashed curves) with the micro-CT image showing the initial condition of the shear simulation. Note that profiles are vertically offset for clearer illustration.

3D shear simulations would mimic at best the deformation processes, but this is currently impossible due to computational limitations. Instead, we built the 2D FDEM model that considers not only the geometry of the experimental specimen but also the initial contact condition on the rough surface. A 2D circular profile at the radius of 5.6 mm, which corresponds to the roughest region (i.e., highest asperities) on the surface, was extracted (Fig. 4a-c). We chose such a profile because the work by Zhao et al. (2018) suggested that this region with the largest roughness plays an important role in controlling the shear strength and fracture development during the experiment. To capture the geometry of the slipping surface, we digitized the top and bottom surfaces before the experiment using a 3D surface scanner (ATOS II by GOM) at a horizontal grid interval of 44 μm. The relative location of the two profiles were adjusted to recreate the initial contact conditions according to the micro-CT image (Fig. 4d-e). We subsampled the profiles to a 0.1 mm nominal grid interval, which was chosen as an acceptable compromise between computation time and accuracy in representing the surface geometry. In addition, to mimic the rotary shear behavior, the two ends of the profiles were extended by 3 mm (i.e., the desired total shear displacement) using the same geometry as their opposite ends to create an effective periodic boundary. These profiles formed the initial shear surfaces of the numerical model (Fig. 4e).

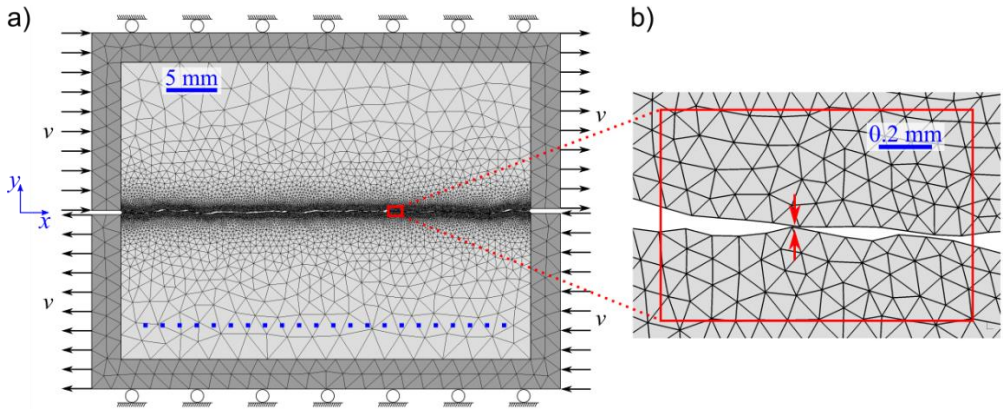

**Fig. 5** (a) Mesh topology and boundary conditions of the shear test simulation. The blue dotted line indicates the location of the virtual measurement line. (b) Zoom in view of the refined mesh at the shear surfaces, and the arrows indicate the smallest gap between top and bottom surfaces.

The bodies of the top and bottom model were 15 mm in thickness, resulting in a total vertical height of 30 mm, similar to the sample used in the laboratory experiment (Fig. 5a). The corners at the ends of the shear surfaces were filleted with a radius of 0.2 mm to avoid stress concentrations that may result in unrealistic damage. To reduce computational time in applying the normal stress during the simulation, the initial vertical distance between the top and bottom semi-sample was adjusted to $2 \times 10^{-6}$ mm (Fig. 5b). Moreover, two rigid boxes were added to simulate the sample holders encasing the two semi-samples. The region of interest (i.e., within 1 mm distance from the shear surface) was discretized with a constant nominal element size of 0.1 mm. The remaining parts of the model were meshed with linearly increasing mesh size as a function of the distance from the shear surface, with the coarsest element size being 3 mm. As a result, the model was meshed into 20,240 triangular elements. These elements were assigned with the calibrated numerical properties (Table 1&2), while the shear boxes had properties of stainless steel (Young's modulus at 200 GPa, density at 8100 kg/m$^3$, and Poisson's ratio at 0.25).

**Table 1** Laboratory measured macromechanical properties (i.e., calibration targets) and emergent properties of the calibrated FDEM model (after Tatone and Grasselli, 2015; Zhao, 2017).

| Properties (unit) | Laboratory measurement | Calibrated FDEM model |
|---|---|---|
| Density (kg·m$^{-3}$) | 1704 | 1704 |
| Young's modulus (GPa) | 15.0 | 15.0 |
| Poisson's ratio (-) | 0.24 | 0.24 |
| Internal friction angle (Degrees) | 23 | 23 |
| Internal cohesion (MPa) | 16.4 | 16.4 |
| Tensile strength (MPa) | 2.6 | 2.7 |
| Uniaxial compressive strength (MPa) | 50.3 | 49.9 |

FDEM models synthesize the macroscopic behavior of materials from the interaction of the micromechanical constituents. The overall deformation and failure behavior of the simulated material are controlled by the combined effect of the input parameters defining the elastic triangular elements and CCEs. As a result, the macroscopic mechanical properties (as listed in Table 1, except for the density that needs no calibration) measured by standard laboratory tests cannot be used directly. Rather, an iterative calibration approach is carried out to obtain input parameters representative of the material, and the laboratory measured properties were used as the calibration targets. In this approach, numerical compressive and tensile strength test models are created and simulated using an initial set of input parameters. The macroscopic mechanical properties and failure patterns are obtained from the simuation and compared against laboratory measurements. In a successful calibration, the numerical model will replicate both the macroscopic mechanical properties measured from the experiments and the overall failure mode of the material. If the simulation result is inadequate, the input parameters are iteratively fine-tuned until the calibration targets are met (Tatone and Grasselli, 2015). The laboratory-measured properties and the emergent macromechanical properties of the calibrated FDEM model are listed in Table 1, and the calibrated FDEM model parameters are listed in Table 2.

**Table 2** Calibrated FDEM model input parameters (after Zhao, 2017).

| Parameter (unit) | Value |
|---|---|
| ***Continuum triangular elements*** | |
| Density, $\rho$ (kg·m$^{-3}$) | 1704 |
| Young's modulus, $E$ (GPa) | 15.6 |
| Poisson's ratio, $\upsilon$ (-) | 0.22 |
| Viscous damping factor, $\alpha$ | 1 |
| ***Cohesive crack elements*** | |
| Internal cohesion, $c$ (MPa) | 17.5 |
| Tensile strength, $\sigma_t$ (MPa) | 2.55 |
| Friction angle, $\phi$ (Degree) | 24.5 |
| Mode I fracture energy, $G_{Ic}$ (J·m$^{-2}$) | 3.8 |
| Mode II fracture energy, $G_{IIc}$ (J·m$^{-2}$) | 90 |
| Fracture penalty, $P_f$ (GPa) | 156 |
| Normal contact penalty, $P_n$ (GPa) | 156 |
| Tangential contact penalty, $P_t$ (GPa) | 156 |

## 2.5 Simulation procedure and boundary conditions

The simulation was computed using the Irazu FDEM software (Geomechanica Inc., 2021) with GPU (graphics processing unit) parallelization. The shear test simulation was conducted in three phases (Table 3). In phase 1, the initial normal stress was applied by compressing the sample at a constant vertical velocity of 0.2 m/s until the vertical stress reaches 2.5 MPa, which corresponds to the initial normal stress condition of the laboratory experiment. In phase 2, the top and bottom boxes were constrained to their vertical position, and the horizontal shear velocity was increased gradually to 0.3 m/s. This transition phase allows the oscillation induced by the instantaneous stop of normal loading to dampen oscillations due to the shear acceleration. In phase 3, the top and bottom boxes were fixed in their vertical positions (i.e., this is a constant normal stiffness shear test) and moved in the horizontal direction at a constant velocity of 0.3 m/s until the desired shear displacement of 3 mm was reached. Note that the loading velocities used in the study are significantly higher (1000 times) than those used in laboratory experiments; however, such a speed has been verified to provide a quasi-static loading condition while allowing a reasonable computation time (Mahabadi, 2012). The model has 26 million simulation time steps, and each step represents a simulation time of $4 \times 10^{-10}$ s.

**Table 3** Simulation phases and boundary conditions applied to the model. The applied velocities in the x ($v_x$) and y ($v_y$) directions, and the resultant shear displacement ($u$) are listed.

| Phase | Simulation steps | $v_x$ (m/s)[1] | $v_y$ (m/s)[2] | $u$ (mm) |
|---|---|---|---|---|
| 1 | 1–66,400 | 0 | 0.1 | 0 |
| 2 | 66,401–964,000 | 0–0.15[3] | 0 | 0–0.02 |
| 3 | 964,000–26,000,000 | 0.15 | 0 | 0.02–3.02 |

[1] Positive (→) on the top box and negative (←) on the bottom box.

[2] Negative (↓) on the top box and positive (↑) on the bottom box.

[3] Linearly interpolated every time step to ramp up the shear velocity gradually.

Normal and shear stresses were measured along a line parallel to the fault and placed 5 mm above the rigid box in the bottom sample (Fig. 5a). This measurement line monitored the stress conditions every 13,000 simulation steps, equivalent to a 200 kHz monitoring rate. The recorded stress values in all elements along the measurement line were averaged to obtain the overall normal stress ($\sigma_n$) and shear stress ($\tau$), which were used to calculate the friction coefficient $\mu = \tau/\sigma_n$, similar to the laboratory-measured apparent friction coefficient.

## 3 Results and data analysis

### 3.1 Shear behavior

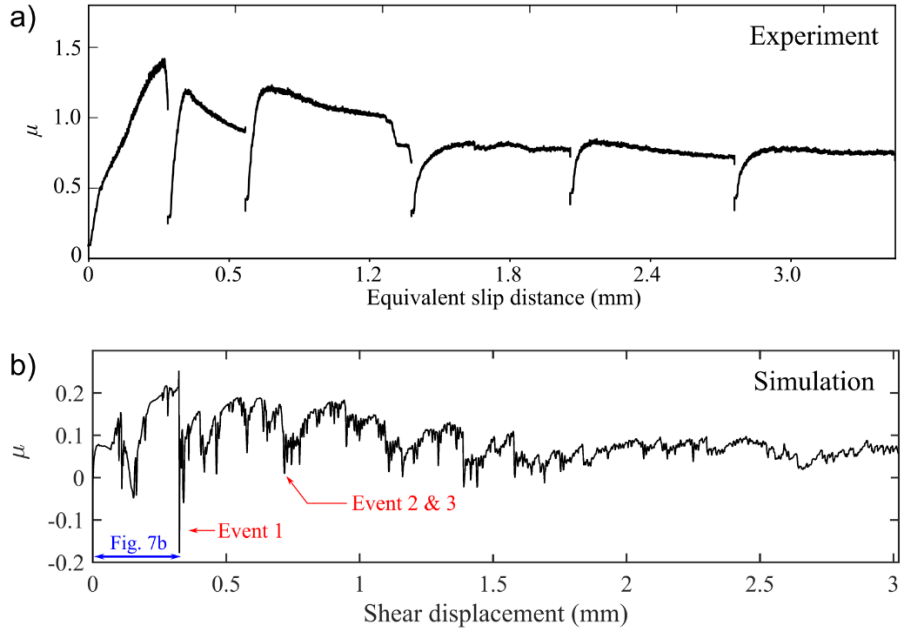

**Fig. 6** Calculated friction coefficient of (a) the laboratory test results plotted as a function of the equivalent slip distance at a radius of 5.6 mm and (b) the numerical simulation. The first ~0.3 mm are detailed in Fig. 7. Red arrows indicate significant drops of frictional resistance associated to seismic events 1, 2, and 3, which are investigated in Section 3.2.

The simulated $\mu$ showed a similar trend with the experimental data. It reached the peak value of 0.22 at a shear displacement of 0.32 mm, followed by a significant drop (Fig. 6b). The simulated $\mu$ experienced many abrupt drops during the slipping process and then stabilized at approximately 0.04 after approximately 1.7 mm of shear displacement. The simulated $\mu$ was significantly lower than the value reported in the laboratory experiment, with many more oscillations.

The simulated stress conditions of the first 0.3 mm showed intriguing similarities to the laboratory experimental data (Fig. 7a&b). In this interval, the shear behavior observed in the experiment can be divided into four stages (Fig. 7a): (I) $\tau$, $\sigma_n$, and the resultant $\mu$ ramped up gradually; (II) $\tau$ experienced a relatively stable stage with minor change, and $\sigma_n$ decreased continuously, causing minor change of $\mu$; (III) $\tau$ and $\sigma_n$ gradually increased to a peak shear stress, and $\mu$ increased to the peak value; and (IV) $\tau$, $\sigma_n$, and $\mu$ dropped rapidly.

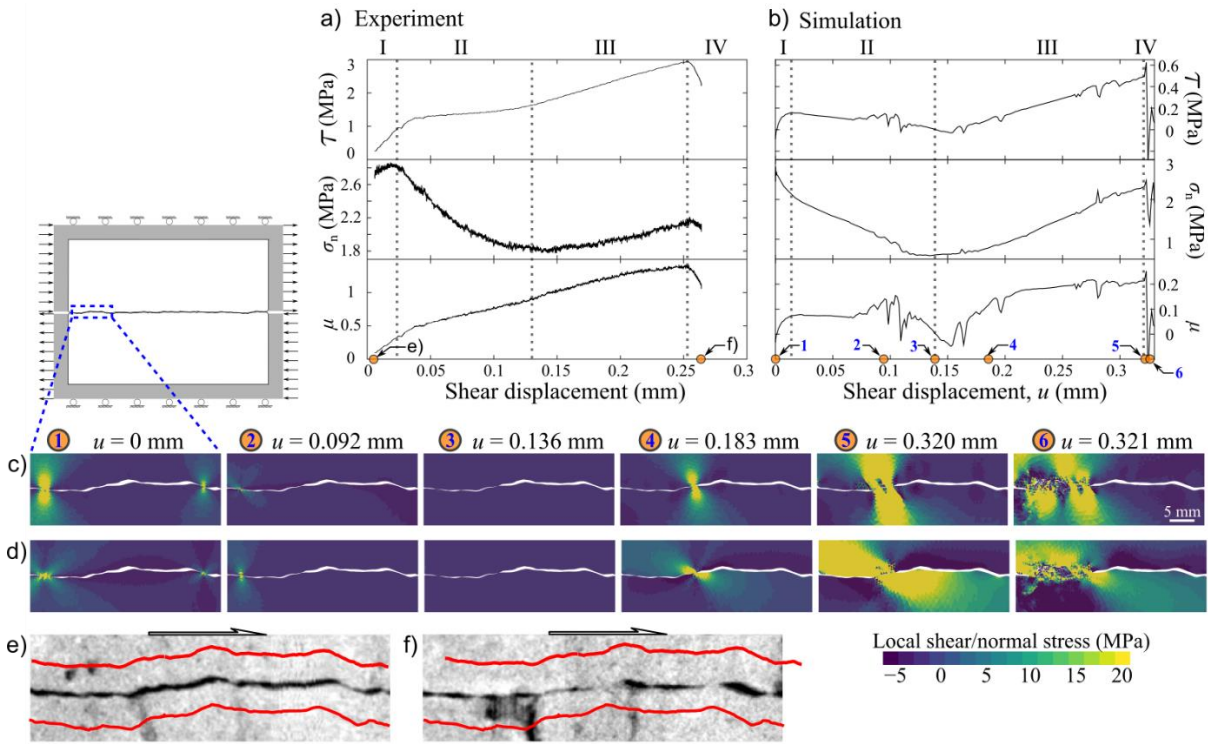

**Fig. 7** Comparison of the overall normal and shear stresses and the friction coefficient of stages I-IV between (a) the experimental data and (b) the simulated data. (c) and (d) are the zoom in views of the local shear and normal stresses, respectively, at the asperity responsible for the stress drop at stage IV. Orange circles numbered 1-6 indicate the horizontal shear displacements ($u$). (e) and (f) are the micro-CT image of the laboratory specimen corresponding to frame 1 and 6 in (c) and (d). The initial surface profiles from the surface scan data (red curves) are placed next to the laboratory fault for comparison.

The numerical simulation qualitatively captured the general trend of these stages (Fig. 7b); however, simulated $\sigma_n$ in stage I decreased gradually, and more oscillations are observed in the curves in the simulated data. To further investigate the mechanisms behind the shear behavior during these stages, we examined the simulated local stress conditions around the asperity whose breakage was responsible for the large and sudden drop of frictional resistance at stage IV (Fig. 7c&d). During stage I, the simulation shows that the shear surface is at the initial contact condition. As the shear displacement increases, the shear stress increases gradually due to frictional resistance of the initial contact area. Note that the numerical model did not capture the minor normal stress increase measured in the experiment at this stage. Such an increase may be related to the interaction of the asperities in the direction perpendicular to shear (i.e., out-of-plane motion) that does not exist in the 2D simulation. During stage II, the top and bottom surfaces adjusted to a more conforming contact, which resulted in the decrease of the shear and normal stress. During stage III, new contact points were established, and asperities engage and interlock, causing the shear stress to increase rapidly reaching the peak shear stress at the end of this stage. Asperities survived and climbed onto each other, causing dilation that increased the normal stress. At stage IV, the highly stressed asperity underwent high-stress concentration and failure, releasing the accumulated strain energy that resulted in the sudden and significant drop of stresses and frictional resistance. The simulated failure pattern, in terms of location and mechanism, resembled the laboratory observation (Fig. 7e&f). More importantly, the numerical model can provide the evolution of surface contacts and stress conditions throughout the shear process.

### 3.2 Progressive damage, gouge formation, and seismic activity

Progressive damage on the shear surface and fault gouge formation was simulated by BCCEs (Fig. 8). The first several BCCEs occurred when the top and bottom semi-sample were loaded with the initial normal stress. Before ~1 mm of shear displacement, the damage was concentrated in the vicinity of the shear surface. After ~1 mm of shear displacement, a number of sub-vertical fractures penetrated the sample body, resembling the fracturing observed in the laboratory (Fig. 9). The distribution of the shear-induced damage was mostly concentrated close to the fault surface and heterogeneously distributed along the fault. Broken asperities formed the gouge layer that accumulated between the semi-samples. As a result, some portions of the bare fracture surface were protected from wearing (Fig. 9a), and this phenomenon is also observed in the laboratory micro-CT image (Fig. 9b).

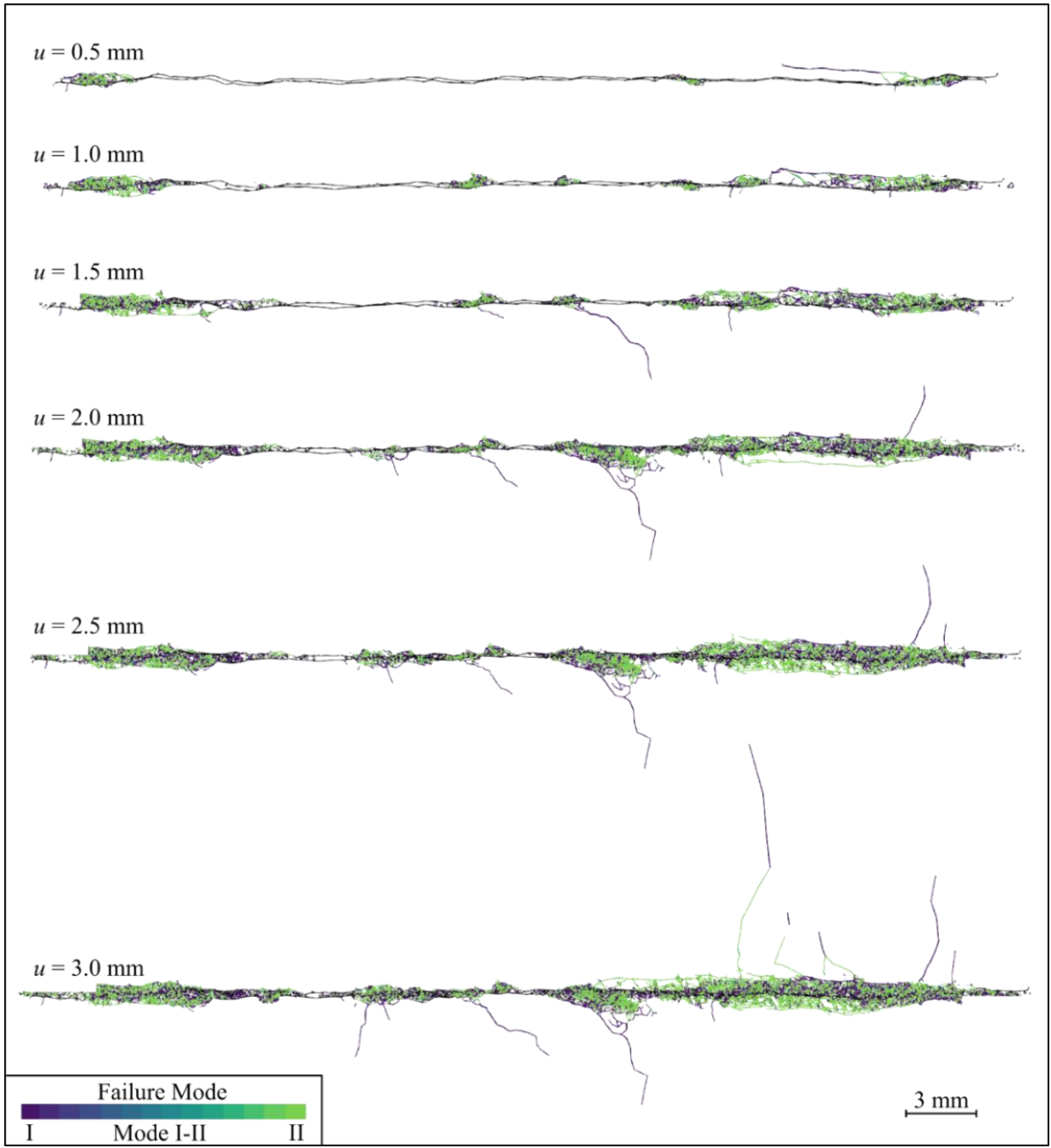

**Fig. 8** Damage of the shear surface and the accumulation of gouge material with increasing shear displacement. Damage is represented by broken cohesive crack elements.

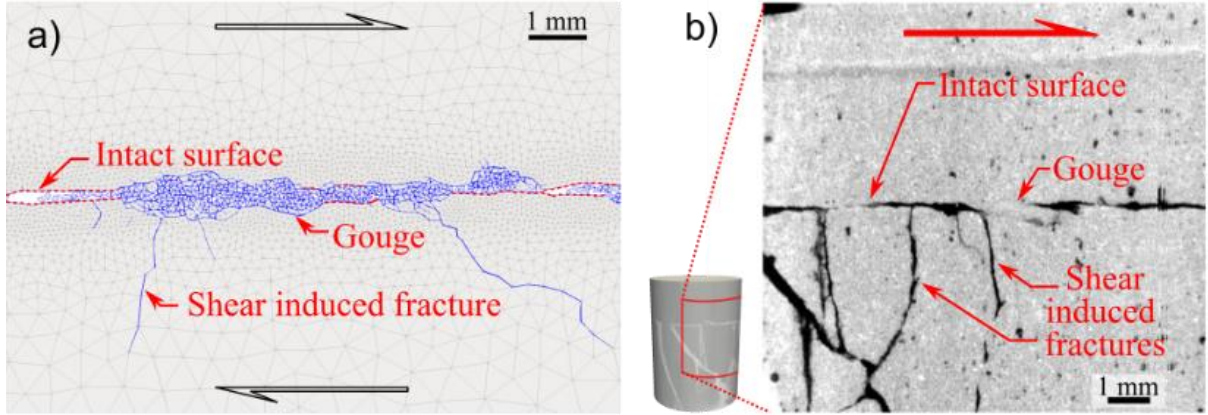

**Fig. 9** (a) Zoom-in view of a portion of the simulated fault surface at 3 mm of slip. The dashed red lines highlight intact fault walls that were not damaged. (b) Zoom-in view of the micro-CT image of a portion of the laboratory fault at a similar location to (a) (adopted from Zhao et al., (2018)).

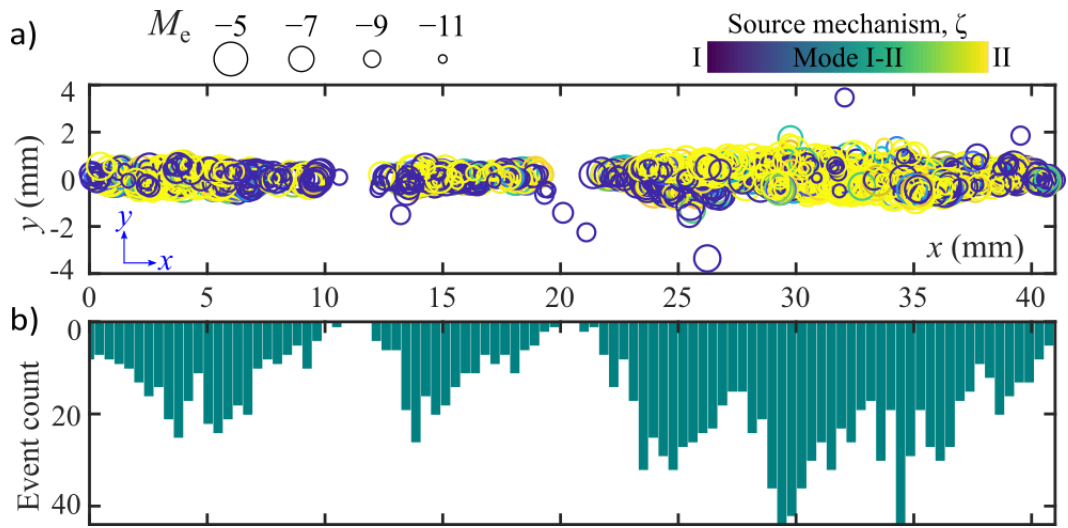

**Fig. 10** Simulated seismic activities. (a) Magnitude, location, and failure mode of the clustered seismic activity. (b) Event count in each bin.

A total of 7,557 CCEs were broken throughout the simulation, and they were clustered into 1,561 seismic events. Most of the BCCEs near the shear surface failed in shear mode (Mode II), and almost all sub-vertical fractures propagated in tensile mode (Mode I). The magnitude of these seismic events ranged between −11.1 and −4.4, with an average magnitude of −7.3. In general, large magnitude events were mostly produced by shear-mode failures, while small magnitude events mostly arose from tensile-mode failures (Fig. 10a). Along the vertical direction ($y$ direction), the spatial distribution of seismic activity coincides with the damage pattern: events were concentrated within ±4 mm of the fault. We divided the horizontal length of the fault, i.e., the x direction, into 100 bins and examined the spatial distribution of the seismic events along such a path (Fig. 10b). Seismic events were distributed heterogeneously along the fault: bins at $x$ = 29.8 mm and 34.9 mm had the largest number of events at 44; bins at $x$ ranging 10.5 to 12.2 mm and 20.5 to 21.0 mm had no seismic events.

Each asperity failure resulted in the sudden and significant drop of frictional resistance and the release of accumulated strain energy. These large magnitude events caused stick-slip-like responses and released high amplitude stress waves propagating across the model. Prior to these dynamic seismic events, their corresponding locations experienced low shear velocity due to the interlocking of asperities and are referred to as interlocking zones (ILZs) in the following discussion. Stress concentrated at ILZs and eventually broke the asperities, releasing the accumulated strain energy (see animated figures Fig. S1 in Supplementary Material for the velocity fields). Three seismic events (Events 1-3, as indicated in Fig. 6) with distinct wave radiation patterns are chosen as examples for further examination. Event 1 at $u$~0.3 mm was related to the most significant friction drop. Events 2 and 3 were two consecutive events that occurred on the slipping surface 12.5 mm apart from each other with a 0.005 ms time delay. We examined the stress field (Fig. 11) and observed that the magnitude of seismic events was directly correlated to the magnitude of the stress concentration at the asperities that failed. We observed that stress concentration at the ILZs reached values as high as the compressive strength of the material, causing compressive failure. Due to interlocking, the non-interlocking regions slightly ahead (with respect to the shear direction) of the ILZs were subjected to significant tensile stress that reached the tensile strength of the material, thus, causing tensile fracturing. By examining the particle velocity field (Fig. 12), we found that prior to the seismic events, the locations of the ILZ were experiencing particle velocities lower than the loading velocity (i.e., < 0.1 m/s). As the seismic events occurred, the source region had particle velocities that were two orders of magnitude higher than that of the ILZs (i.e., > 10 m/s). Interestingly, considering that P- and S-wave velocities are 2967 m/s and 1884 m/s, respectively, Event 3 occurred right after the arrival of the P-wave induced by Event 2, but prior to the arrival of the S-wave. Therefore, Event 3 may have been triggered by the stress perturbation from Event 2.

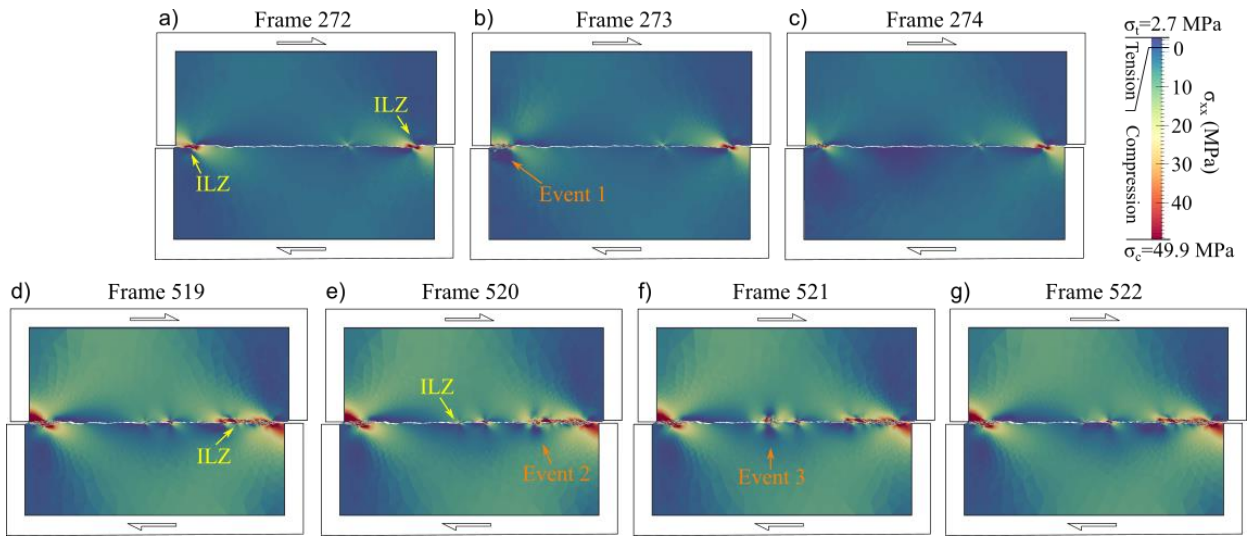

**Fig. 11** Output frames of the numerical model showing the horizontal stress $\sigma_{xx}$ at (a-c) Event 1 and (d-g) Events 2 and 3. Interlocking zones (ILZs) that are related to the selected seismic events are highlighted by the yellow arrows. Note that the simulation time interval between two frames is 0.052 ms.

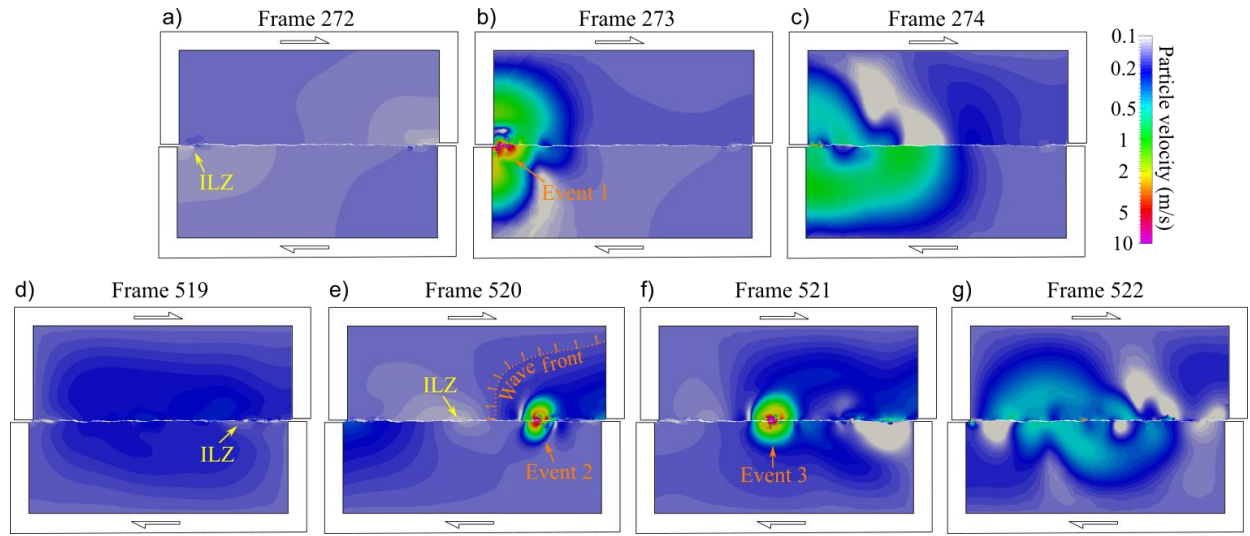

**Fig. 12** Output frames of the numerical model showing the particle velocity at (a-c) Event 1 and (d-g) Events 2 and 3. ILZs are highlighted by the yellow arrows and the P-wave wavefront of Event 2 is labelled. Note the color map is in log scale.

## 4 Discussion

The FDEM numerical model qualitatively captured the mechanical behavior observed in the laboratory experiments, highlighting the dominant role of surface roughness on the shear behavior of rocks at low-stress conditions. Both the laboratory experiment and the numerical simulation show a slip weakening behavior where the friction coefficient ramps up to the peak value and then decreases to a residual value. In the numerical simulation, the shear stress and friction reached a steady-state and residual value around ~1.7 mm of total displacement, in agreement with the laboratory value (Zhao et al., 2018).

During the first ~0.3 mm of shear displacement, the discrepancies between the experimental and simulation results in stage I and the additional stress oscillations in the simulation results may be because the 2D model was not able to capture 3D asperity interactions. However, the overall variation trend of stresses and the damage pattern on the shear surface showed close similarities, suggesting that the 2D profile that we used is a proxy for the laboratory specimen. This also supports our previous interpretation that the highest asperity was responsible for the formation of the large secondary sub-vertical fractures and the associated sudden drop in shear resistance (Zhao et al., 2018). These results suggest that our numerical technique, which uses a combination of surface scanning, X-ray micro-CT imaging, and FDEM modelling, represents a promising approach to simulate realistic fault behavior. Our simulation provides the continuous evolution of contacts on the shear surface and the stress conditions that complement the laboratory observations in achieving a better comprehension of how the interaction between asperities controls the stress conditions and damage patterns in faults.

During the shear process, asperities interact in various modes including climbing onto each other, interlocking, and breaking (Scholz, 1990). Our experimental and numerical results show that such interactions directly influenced the stress conditions and damage patterns. When the slip displacement is small ($u < 1.5$ mm), weak asperities (i.e., millimetric scale unevenness) controlled the frictional behavior, creating gouge material. These observations agree with the laboratory observations on the post-mortem sample and suggest the importance of surface roughness in controlling the formation of the gouge layer. As the slip displacement increases ($u > 1.5$ mm), the large-scale roughness of the shear surface (i.e., centimetric scale waviness) becomes important to the shear behavior and damage pattern. Large scale waviness causes high stress concentration through interlocking and climbing and may cause sub-vertical secondary fractures.

The damage and seismic event distributions are closely related to the stress heterogeneity on the shear surface caused by the surface roughness. Depending on the geometry of the asperity, the stress concentration at the ILZs could reach the compressive strength of the material, causing compressive failure. This mechanism creates gouge material in the vicinity of the shear surface.

On the other hand, the areas ahead of the ILZs experience tensile stress up to the tensile strength of the material, thus, creating tensile fractures. This mechanism creates large sub-vertical secondary fractures. Breakage of strong asperities release the accumulated strain energy in the whole model, causing an overall shear stress drop, giving a stick-slip-like shear behavior. Such a lock-and-fail mechanism is recently found to be the key process of stick-slip behavior of bare surfaces (Chen et al., 2020; Morad et al., 2022). Note that the overall shear loading in our model is considered quasi-static, but the local seismic events are dynamic activities with particle velocity more than 100 times the quasi-static loading velocity. This suggests that on a rough shear surface, quasi-static shear consists of numerous heterogeneously distributed local dynamic seismic activities, and this process may complicate the slip process on rough faults and the estimation of the energy budget (Tinti et al., 2005).

Observations on Events 2 and 3 suggest that the stress perturbation from asperities breakage may trigger events on adjacent interlocking zones. From an earthquake perspective, there are two possible mechanisms that may trigger seismic events in the near field: (1) static stress redistribution (e.g., King et al., 1994; Toda et al., 1998) and (2) dynamic stress wave perturbation (e.g., Kilb et al., 2000; Gomberg et al., 2001). In our simulation, the modeled body did not slip as a rigid body, rather, the slipping consisted of pulses of local movements, accompanied by numerous continuously changing of contacts and asperities breakages. When the asperity associated to Event 1 breaks, the dynamic stress perturbation was damped out, and the static stress concentration is transferred to nearby asperities, which eventually caused failure of other asperities. On the other hand, Events 2 and 3 showed a more interesting correlation. Event 3 occurred between the arrival times of P- and S-waves from Event 2. Within this time window, stress redistribution had not reached a steady state, suggesting that the perturbation of the dynamic stress wave radiated from Event 2 may have triggered Event 3. These results imply that static stress transfer and dynamic stress perturbation triggering may occur on the same fault and contribute to the movement of fault slip. However, due to the limitation of the model output frequency and post-processing method, the triggering is not conclusive, Event 2 and 3 may have been independent seismic events occurred in a narrow time window, and more investigation is needed in future research.

The numerical simulation has the advantage of continuously modeling the fault shear process, fault surface damage, and associated stress conditions. However, the simulated sample experienced more damage than the laboratory sample, which is probably related to the limitation of 2D simulations not accounting for the motion in the third dimension. For the same reason, the simulated stresses suffered significant fluctuations, and the friction coefficient was much lower than the experimental measurement, which is a common limitation of 2D simulations. The laboratory experiment by Frye and Marone (2002) and the numerical simulation by Hazzard and Mair (2003) demonstrated that 2D numerical models exhibit friction values notably lower than 3D models and suffer from greater stress fluctuations due to the lack of particle motion in the third dimension. In addition, we meshed the shear surface at a relatively high resolution (0.1 mm), resulting in a large number of asperities at various sizes. Hence, the interlocking and breakage of these asperities caused stress oscillations (i.e., microseismic events). Even though we qualitatively captured the shear behavior that matches the laboratory measurements, to fully capture the shear behavior of the rotary shear experiment, a 3D model capturing the surface geometry and asperity interaction on the entire shear surface will be required.

## 5 Conclusion

In this study, we used a carefully built and calibrated FDEM numerical model to simulate a laboratory shear experiment. We introduced a new clustering algorithm to improve the understanding of the simulated fracturing and associated seismic events. The model was able to qualitatively capture the frictional behavior observed in the laboratory experiment, providing the missing information in the experimental observation regarding the continuous variation of stresses and the progressive evolution on the shear surfaces.

Our numerical model matches the experimental results particularly well at the beginning of the shear deformation (~0.3 mm). We were able to identify similar stress variation trends and damage patterns. The simulation results provided detailed evolution processes of the contacts on the shear surface and the local stress conditions, which are not available in experimental observations. Combining the numerical and experimental results, we conclude that interlocking of asperities can cause compressive stress concentration on the front side (i.e., facing the shear direction) of the asperity, which could induce compressive failure (e.g., crushing) near the shear surface; on the other hand, tensile stress concentration is generated on the leeward side of the asperity, which could cause sub-vertical tensile fractures that could propagate into the host rock. Progressive surface damage and the associated microseismic events occur at the locations of asperity interactions and is highly heterogeneous. Several locations experienced no damage even after large shear displacement, these locations are either not in contact or were protected by gouge materials.

As a result of the interlocking and breakdown of asperities, local dynamic failure events occur, even though the overall loading is quasistatic. These events are considered microseismic events, and their magnitudes range between −11.1 and −4.4. Strain energy stored in the medium was released during these events, causing dynamic perturbation to the overall stress condition, and the particle velocity in the source reached > 10 m/s, two orders of magnitude larger than the surrounding regions. This high amplitude stress perturbation could even trigger the failure of adjacent critically stressed asperities.

Both the numerical model and the experiment suggested the importance of shear surface roughness in controlling slip behavior, and we were able to explain the laboratory observations with the help of numerical results. Shear surface evolution is a complicated process that involves frictional sliding, fracturing, gouge comminution, and seismicity. The high degree of agreement between simulation and experiment data leads to a promising future of predicting fault behavior through, laboratory testing, surface characterization, and numerical simulations. These results improved the understanding of shear behavior and demonstrated that micromechanical based numerical simulation is a capable approach to study fault mechanics.

**Declaration of competing interest**

The authors declare that they have no known competing financial interests or personal relationships that could have appeared to influence the work reported in this paper.

**Acknowledgements**

Q. Zhao is supported by the FCE Start-up Fund for New Recruits at the Hong Kong Polytechnic University (Project ID P0034042) and the Early Career Scheme of the Research Grants Council of the Hong Kong Special Administrative Region, China (Project No. PolyU 25220021). This work has also been supported through the NSERC Discovery Grants 341275, CFILOF Grant 18285, Carbon Management Canada (CMC), and NSERC/Energi Simulation Industrial Research Chair Program. The authors would like to thank Geomechanica Inc. for providing the Irazu FDEM simulation software. Q. Zhao would like to thank Dr. Andrea Lisjak and Dr. Bin Chen for discussions and suggestions. The authors appreciate the constructive suggestions and comments from the editor and the reviewers.